\documentclass[usenatbib]{mn2e}
\usepackage[dvips]{graphicx}
\usepackage{amssymb}
\usepackage{txfonts}
\usepackage{colordvi}
\usepackage{color}
\usepackage{url}

\newcommand{\msun}{{\rm M}_{\sun}}

\newcommand{\g}{$\gamma$}

\newbox\grsign \setbox\grsign=\hbox{$>$} \newdimen\grdimen \grdimen=\ht\grsign
\newbox\simlessbox \newbox\simgreatbox \newbox\simpropbox
\setbox\simgreatbox=\hbox{\raise.5ex\hbox{$>$}\llap
     {\lower.5ex\hbox{$\sim$}}}\ht1=\grdimen\dp1=0pt
\setbox\simlessbox=\hbox{\raise.5ex\hbox{$<$}\llap
     {\lower.5ex\hbox{$\sim$}}}\ht2=\grdimen\dp2=0pt
\setbox\simpropbox=\hbox{\raise.5ex\hbox{$\propto$}\llap
     {\lower.5ex\hbox{$\sim$}}}\ht2=\grdimen\dp2=0pt
\def\ga{\mathrel{\copy\simgreatbox}}
\def\la{\mathrel{\copy\simlessbox}}

\title[Recollimation shocks in high-mass X-ray binaries]{Formation of recollimation shocks in jets of high-mass\\ X-ray binaries}

\author[D.S. Yoon et al.]
{DooSoo Yoon,$^{1,2}$ Andrzej A. Zdziarski$^3$ and Sebastian Heinz$^1$\\
  $^1$Department of Astronomy, University of Wisconsin-Madison, Madison, WI, USA\\
$^2$Shanghai Astronomical Observatory, Chinese Academy of Sciences, 80 Nandan Road, Shanghai 200030, China\\
  $^3$Centrum Astronomiczne im.\ M. Kopernika, Bartycka 18, PL-00-716
  Warszawa, Poland }

\date{Accepted 2015 December 16.  Received 2015 November 26; in original form 2015 August 19}

\pagerange{\pageref{firstpage}--\pageref{lastpage}}
\pubyear{2015}

\begin{document}

\maketitle

\label{firstpage}

\begin{abstract}
  We study conditions for formation of recollimation shocks in jets
  interacting with stellar winds in high-mass X-ray binaries. We show
  the existence of a critical jet power, dependent on the wind rate
  and velocity and the jet velocity, above which a recollimation shock
  is not formed. For jet powers below critical, we derive the location
  of the shock. We show that surface shocks may still exist above the
  critical power, but only occupy a small volume of the jet and do not
  significantly alter the jet opening angle. We test these prediction
  by 3-D numerical simulations, which confirm the existence and the
  value of the critical power. We apply our results to Cyg X-1 and Cyg
  X-3.
\end{abstract}

\section{Introduction}
\label{intro}

Interaction of an extragalactic jet with a stellar wind of a massive
star passing through the jet was considered by \citet{bp97}. As the
shock-formation condition, they used the balance of the ram pressures
of the forward motion of the jet matter and that of a radial stellar
wind. This appears to be the proper approximate condition if the
velocity of the star within the jet is much less than the wind
velocity. See also \citet{komissarov94} for the case of jet
interaction with winds of low-mass stars.

The same condition was then used by \citet*{dch10} for formation of a
recollimation/reconfinement shock in a jet originating from a compact
object and interacting with the stellar wind of a massive donor in a
binary system. The condition formulated by \citet{dch10} was then
applied to Cyg X-3 by \citet{z12}, who concluded from it that a
recollimation shock can be formed only if the initial jet half-opening
angle, $\alpha_0$, is rather large, $\ga\! 30\degr$.

However, the wind-jet collision in a binary system is clearly not
head-on. Close to the binary plane, the wind velocity is parallel to
the plane, and the jet is launched away from it. In the case of a jet
perpendicular to the binary plane, see Fig.\ \ref{fig:sketch}, the jet
and wind velocities are close to perpendicular along the binary plane,
and only a velocity component related to the jet opening angle is
collinear with the wind.

This effect was taken into account in analytical estimates of
\citet{yh15}, hereafter YH15 (see \citealt{perucho08} and
\citealt*{perucho10} for previous studies of formation of shocks in
X-ray binary jets). As the condition of the ram pressure of the jet
being in equilibrium with the wind ram pressure, YH15 took only the
jet velocity component collinear with the wind direction in the
vicinity of the binary plane, i.e., $v_{\rm j}\sin\alpha_0$, where
$v_{\rm j}$ is the jet velocity and $\alpha_0$ is the initial opening
angle of the jet. This is applicable only if the recollimation shock
takes place at a distance from the jet origin of $z\ll a$, where $a$
is the separation between the binary components. They derived a
formula for the location of the recollimation shock in this
approximation. YH15 also performed numerical simulations of the
jet-wind interaction and found their formula agrees relatively well
with the simulation results in the case of a moderate jet power,
$10^{37}$ erg s$^{-1}$, interacting with a strong wind with a high
mass loss rate of $\dot M_{\rm w}=10^{-5}\msun$ yr$^{-1}$ and a high
wind velocity of $v_{\rm w}=2.5\times 10^8$ cm s$^{-1}$. However, the
assumption that the shock takes place at $z \ll a$ is not applicable
for a sufficiently strong jet.

Here, we generalize the calculation of YH15 to the case of an
arbitrary jet power. We concentrate on the case of systems in
which the jet and orbital axes are aligned; see, e.g., \citet{sm12} for a
discussion of conditions for the spin-orbit alignment. We test the
calculations by means of numerical simulations. We apply our results
to jets of the high-mass X-ray binaries Cyg X-1 and Cyg X-3.

\section{Formation of recollimation shocks}
\label{formation}

\subsection{The setup of the problem}

\begin{figure}
    \centering
    \includegraphics[width=\columnwidth]{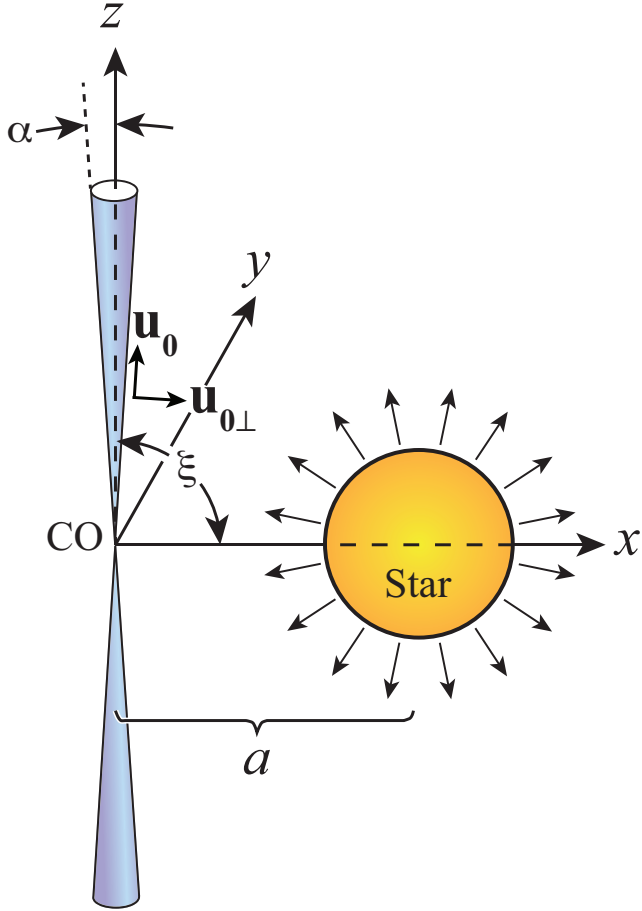}
    \caption{A sketch of a jet originating from a compact object (CO), such as a neutron star or a black hole, and a stellar wind. The jet direction is at an angle, $\xi$, with respect to the direction from the CO to the star. In the case considered here, the jet is along the $z$-direction, and thus $\xi=90\degr$. The jet's half-opening angle is $\alpha$ and the separation is $a$. The star, located at $x_*=a$, emits an isotropic and uniform stellar wind. We also show a unit vector along the contact discontinuity, $\mathbf{u_0}$, and one perpendicular to it, $\mathbf{u_{0\perp}}$.
} 
\label{fig:sketch}
\end{figure}

Suppose the jet is expanding with an initial opening angle $\alpha_{0}$. The internal pressure, $p_{\rm j,i}\propto z^{-2\gamma}$ (where $\gamma$ is the jet's adiabatic index), of the jet drops more rapidly than the jet's lateral ram pressure, $p_{{\rm j},x} \propto z^{-2}$, and the lateral expansion of the jet will become supersonic. Whenever the internal pressure of the jet drops below the total ambient pressure, a shock will be driven laterally into the jet in order to bring the internal pressure into equilibrium with the bow shock.

Such a shock will necessarily reduce the lateral expansion velocity of
the jet (using the free energy in the jet's expansion to increase the
internal energy of the jet fluid), and may thus be called a
``recollimation'' shock. Below, we will distinguish ``strong
recollimation shocks'' from ``weak'' or ``surface shocks''. A strong
recollimation shock will convert {\em most or all} of the lateral
expansion energy (a fraction $\sim \sin^2{\alpha_{0}}$ of the total
kinetic energy of the jet) into particle heating and/or acceleration, with important observational consequences for, e.g., the
high-energy emission from particles accelerated in this shock. A
surface shock will reduce the opening angle by a small amount,
$\Delta \alpha \ll \alpha$, sufficient to raise the pressure to the
ambient value, but dissipating only a small fraction of the jet's
lateral kinetic energy.

To estimate the change in opening angle of the jet, we consider ram
pressure balance of a supersonic spherical wind colliding with a
supersonic conical jet. The collision will take place on the side of
the jet exposed to the wind. At a given point along the jet, this can
be approximated as two streams colliding at some angle. This collision
will lead to formation of two shocks -- one into the wind (which we
will refer to as the bow shock) and one into the jet (the
recollimation shock).

We define the coordinate system with the $x$-$y$ plane coinciding with
the binary plane, and the $z$ axis perpendicular to it, see Fig.\
\ref{fig:sketch} and, for the sake of simplicity, analyse the shock
propeties in the $x$-$z$ plane (where the bow shock and thus the
recollimation shock will be strongest). We place the star centre at
$x_*=+a$ and the jet origin at $x=0$.  Following YH15, we perform all
calculations in the non-relativistic approximation.

The contact discontinuity will be parallel to the post-shock
velocities of the jet and the wind. In the frame of the contact
discontinuity, ram pressure balance takes the form of, e.g., eq.\ (11)
of \citet{ns09},
\begin{equation}
  \rho_{\rm w} v_{\rm w\perp}^2=\rho_{\rm j} v_{\rm j\perp}^2,
  \label{ram_eq}
\end{equation}
where $\rho_{\rm w}$, $\rho_{\rm j}$ are the densities of the wind and
the jet, respectively, and $v_{\rm w\perp}$ and $v_{\rm j\perp}$ are
the wind and jet velocity components perpendicular to the contact
discontinuity, respectively. We can define a unit vector along the
contact discontinuity and a vector perpendicular to it as
\begin{eqnarray}
  \mathbf{u_0} & = & \left(\sin{\alpha'},0,\cos{\alpha'}\right) = 
  \left(x_0,0, \sqrt{1- x_0^2}\right), \nonumber \\
  \mathbf{u_{0\perp}} & = &
  \left(-\cos{\alpha'},0,\sin{\alpha'}\right) =
  \left(-\sqrt{1- x_0^2},0,x_0\right),
  \label{vectors}
\end{eqnarray}
which defines $x_0$ and the inclination angle $\alpha'$ of the contact
discontinuity relative to the initial jet axis. Then
$v_{\rm w\perp} = \mathbf{v_{\bf w}} \cdot \mathbf{u_{0\perp}}$ and
$v_{\rm j\perp}=\mathbf{v_{\bf j}}\cdot \mathbf{u_{0\perp}}$. We can
then solve equation (\ref{ram_eq}) for $x_0(z)$. The trajectory
assuming the post-shock jet fluid follows the contact discontinuity is
then given by the solution of
${\rm d}x/{\rm d}z=\tan{\alpha'}=x_0/\sqrt{1-x_0^2}$.  We note that we
do not impose conservation of energy and momentum, assuming that
energy can be dissipated and radiated and the wind flow can continue
around the jet. Also, the above condition is not satisfied close to
the jet origin because the perpendicular velocity components are not
supersonic. We neglect this effect since we are concerned with regions
where the velocity components are supersonic.

The jet is initially freely expanding and conical with the initial
semi-opening angle, $\alpha_0$, related to its initial Mach number,
$\mathcal{M}_{\rm j}$ (YH15). The lateral ram pressure of the jet,
i.e., along the $x$ direction, is
\begin{equation}
p_{{\rm j,}x}=\rho_{\rm j}\tan^2 \alpha_0 v_{\rm j}^2={P_{\rm j}\over \upi z^2 v_{\rm j}},
\label{pj1}
\end{equation}
where we assumed $v_{\rm j}$ to be constant. In the second equality,
we have expressed the pressure in terms of the kinetic power of the
jet+counterjet in the rest mass motion,
$P_{\rm j}=\upi \rho_{\rm j}v_{\rm j}^3 z^2 \tan^2 \alpha_0$. Then,
$p_{{\rm j,}x}$ is independent of the jet opening angle. Note that we
use the power of both jets here, which is of interest when considering
the energy budget of the system, in particular comparing the jet and
accretion powers. The same convention was used by YH15.

On the other hand, the ram pressure of the wind in its motion
component along the binary plane is
\begin{equation}
  p_{{\rm w,}x}=\rho_{\rm w}v_{{\rm w,}x}^2 = 
  {\dot M_{\rm w} v_{\rm w}\over 4\upi a^2}\left(a^2\over a^2+z^2\right)^2,
  \label{pw}
\end{equation}
where $\dot M_{\rm w}$ is the mass-loss rate in the wind and $v_{\rm
  w}$ is its velocity, assumed constant along the jet. The second
factor on the right-hand side appears due to both the wind density and
$v_{{\rm w,}x}^2/v_{\rm w}^2$ decreasing as $a^2/(a^2+z^2)$. This
factor is not present in eq.\ (8) of YH15 because they consider only
shock formation close to the binary plane. However, as we show below,
recollimation shocks in jets of HMXBs can take place at distances
comparable to the binary separation, or not happen at all, in which
cases that term is important. 

We solve for the inclination angle of the contact discontinuity:
\begin{equation}
  \tan{\alpha'} = \tan{\alpha_{0}} \frac{ \sqrt{4P_{\rm
  j}/\left(\dot{M}_{\rm w}v_{\rm
    w}v_{\rm j}\right)}  \left(a^2 + z^2\right) -
    az} {\sqrt{4P_{\rm j}/\left(\dot{M}_{\rm w}v_{\rm w}v_{\rm
    j}\right)} \left(a^2 +
      z^2\right) + z^2\tan{\alpha_{0}}},
  \label{eq:tana}
\end{equation}
with a deflection angle at the leading edge of the jet of
$\Delta \alpha = \alpha_0-\alpha'$.

\subsection{Strong recollimation shocks}

The formation of a strong recollimation shock corresponds to the
direction defined by equation (\ref{ram_eq}) being substantially
different from the direction of the jet boundary facing the star,
i.e., $\Delta\alpha \geq \alpha$.  The limiting case for the formation
of a strong shock is thus $\alpha'=x_0(z)=0$, which corresponds to the
treatment given in section 3.2 of YH15.

Equating the pressures leads to a quadratic equation in $(z/a)^2$. The
equation has real positive solutions for the jet kinetic power
satisfying
\begin{equation}
  P_{\rm j}\leq P_{\rm cr} \equiv {1\over 16}\dot M_{\rm w}v_{\rm w}v_{\rm j}.
  \label{condition}
\end{equation}
Note that $P_{\rm cr}$ is independent of the jet opening angle. 

More generally, in the case of jets inclined with respect to the
orbit, it is straightforward to show that the shock formation
criterion is
\begin{equation}
  P_{\rm j} \leq P_{\rm cr} \left(\frac{1 + \cos{\xi}}{\sin{\xi}}\right)^2,
\end{equation}
where $\xi$ is the inclination angle between the jet and the orbital
separation vector, ${\mathbf a}$. Note that the critical power is
different for approaching and receding half of an inclined jet, as the
sign of $\cos{\xi}$ changes from positive to negative as a jet tilts
further than $90^{\circ}$ away from the companion, because the wind
ram pressure decreases and lower powers are sufficient to suppress the
formation of a recollimation shock.

If the condition in equation (\ref{condition}) is satisfied, the jet
lateral ram pressure, decreasing as $z^{-2}$, becomes equal to the
corresponding wind ram pressure at some point $z_{1}$ along the jet,
and a strong recollimation shock forms. Thus, the lower solution
corresponds to the condition of shock formation. The shock location is
then given by
\begin{eqnarray}
  \lefteqn{
  \left(z_1\over a\right)^2=\delta-1-\sqrt{\delta^2-2\delta},\quad
  \delta\equiv {\dot M_{\rm w}v_{\rm w}v_{\rm j}\over 8 P_{\rm j}}\geq
  2,
  \label{solution}}\\
  \lefteqn{
  \qquad \simeq 1/(2\delta),\quad \delta\gg 1,\label{approx}}
\end{eqnarray}
where the approximate equation (\ref{approx}) corresponds to the
solution given by YH15 in their eq.\ (9). At the critical point of
$P_{\rm j}=P_{\rm cr}$, there is one solution at $z_1=a$. Thus, under
the adopted approximations, a strong recollimation shock never forms
at $z>a$. If $P_{\rm j}> P_{\rm cr}$, the jet lateral ram pressure
will be always greater than that of the wind.

Once the shock takes place, the jet is no longer conical and freely
expanding [such that equation (\ref{eq:tana}) no longer applies], but instead
its thickness follows from pressure equilibrium between the jet and
the wind bow shock, see section 4.1 of YH15. However, equations
(\ref{pj1}--\ref{pw}) still remain valid. Thus, the jet becomes again
freely expanding above $z_2$ given by the upper solution of the
quadratic equation obtained by equating these two equations,
\begin{equation}
  \left(z_2\over a\right)^2=\delta-1+\sqrt{\delta^2-2\delta},\quad \delta\geq 2.
\label{upper}
\end{equation}

\subsection{Surface shocks}

Even if the strong shock criterion is not satisfied, a shock may still
form if the internal pressure of the jet drops below the bow shock
pressure. In that case, the maximum deflection angle, $\Delta \alpha$,
from equation (\ref{eq:tana}) is small compared to $\alpha$.  While
such a surface shock will necessarily be highly oblique and weak
compared to a recollimation shock, it is important to note that it may
still be a strong shock in the sense of having a large Mach number and
changing the entropy of the gas.

In the limit of $\Delta\alpha \ll \alpha_0$, the location of the maximum
relative dynamic pressure, where $\Delta \alpha$ is maximized, is
\begin{equation}
  z_{\rm maz} \approx \frac{a}{\cos{\alpha_{0}}} \left(1 +
      \sin{\alpha_{0}}\right) \simeq a,
\end{equation}
and, to the lowest order, the maximum deflection angle becomes
\begin{equation}
  \Delta\alpha \approx - \alpha_{0} \sqrt{\frac{P_{\rm cr}}{P_{\rm j}}},
\end{equation}
valid in the regime $P_{\rm j} \gg \dot{M}v_{\rm w}v_{\rm j}$ (and
assuming the small angle approximation for $\alpha_{0}$).

\begin{figure*}
    \centering
    \includegraphics[width=15.2cm]{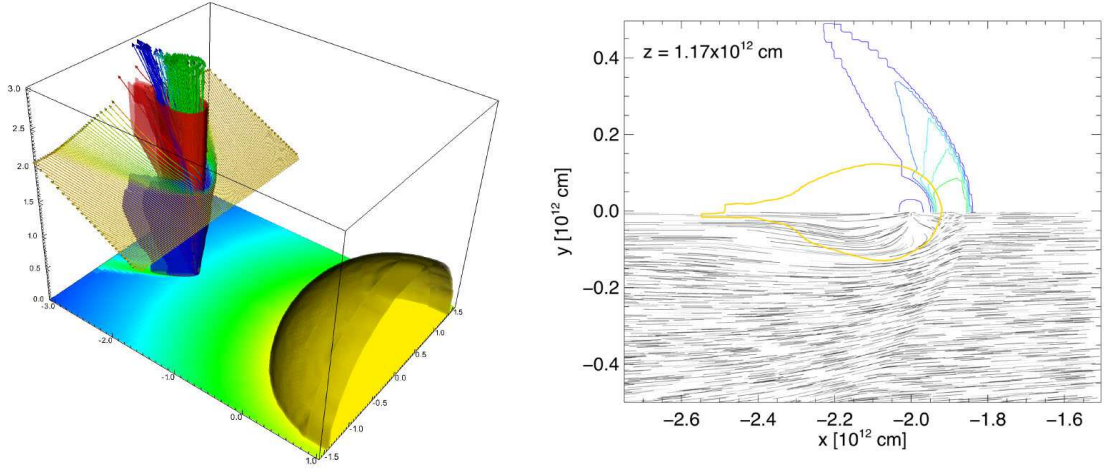}
    \caption{Left panel: Visualization of the P3e37 simulation. {\em
        Red contour surface}: section of the jet boundary; {\em blue
        contour surface:} section of the bow shock; {\em yellow
        contour surface:} surface of the star; {\em bottom surface:}
      slice through the temperature of the stellar wind and bow shock;
      {\em vectors:} stream lines sampling the jet and the stellar
      wind/bow shock. The axes are in units of $10^{12}\,{\rm cm}$.
      Right panel: slice through P3e37 simulation at
      $z=1.17\times 10^{12}\,{\rm cm}$ showing the structure of the
      bow shock and the formation of a surface shock on the leading
      edge of the jet; {\em top part:} five pressure contours linearly
      spaced between 13 and 80 per cent of the theoretical stagnation
      point pressure of the bow shock; overlaid in yellow is the
      approximate location of the contact discontinuity; {\em bottom
        part:} 2-D velocity streamlines in the $x$-$y$ plane showing
      the deflection of the wind in the bow shock and the acceleration
      of jet material away from the stagnation point behind the
      surface shock and in the expansion wave in the down-wind region
      of the jet.}
    \label{fig:render}
\end{figure*}

The pressure behind such a surface shock is still given by the bow
shock pressure, $p_{\rm shock} \sim \dot{M}v_{\rm w}^2/(16\upi a^2)$ (the
same as in the case of a strong recollimation shock). The volume
occupied by the shock is roughly
\begin{equation}
  V_{\rm shock} \sim \frac{2\pi a^3}{3}  \alpha_{0}\Delta \alpha \simeq
  \frac{2\pi}{3} \alpha_{0}^2 a^3 \sqrt{\frac{P_{\rm cr}}{P_{\rm j}}}.
\end{equation}
This situation is illustrated in Fig.~\ref{fig:render}, which shows
the structure of the bow shock and the moderate oblique shock driven
into the jet.

Since we would expect the non-thermal emission from such a shock to be
proportional to the volume of the shock and some function of the
pressure, the total non-thermal emission from the shock should be
proportional to $\sqrt{\dot{M}v_{\rm w}v_{\rm j}/P_{\rm j}}$, i.e.,
it decreases with the square root of the jet power for super-critical jets
that do not form strong recollimation shocks and for otherwise identical
jet and wind parameters.

\section{Comparison with simulations}
\label{simulations}

\begin{figure*}
   \centering
   \includegraphics[width=1\textwidth]{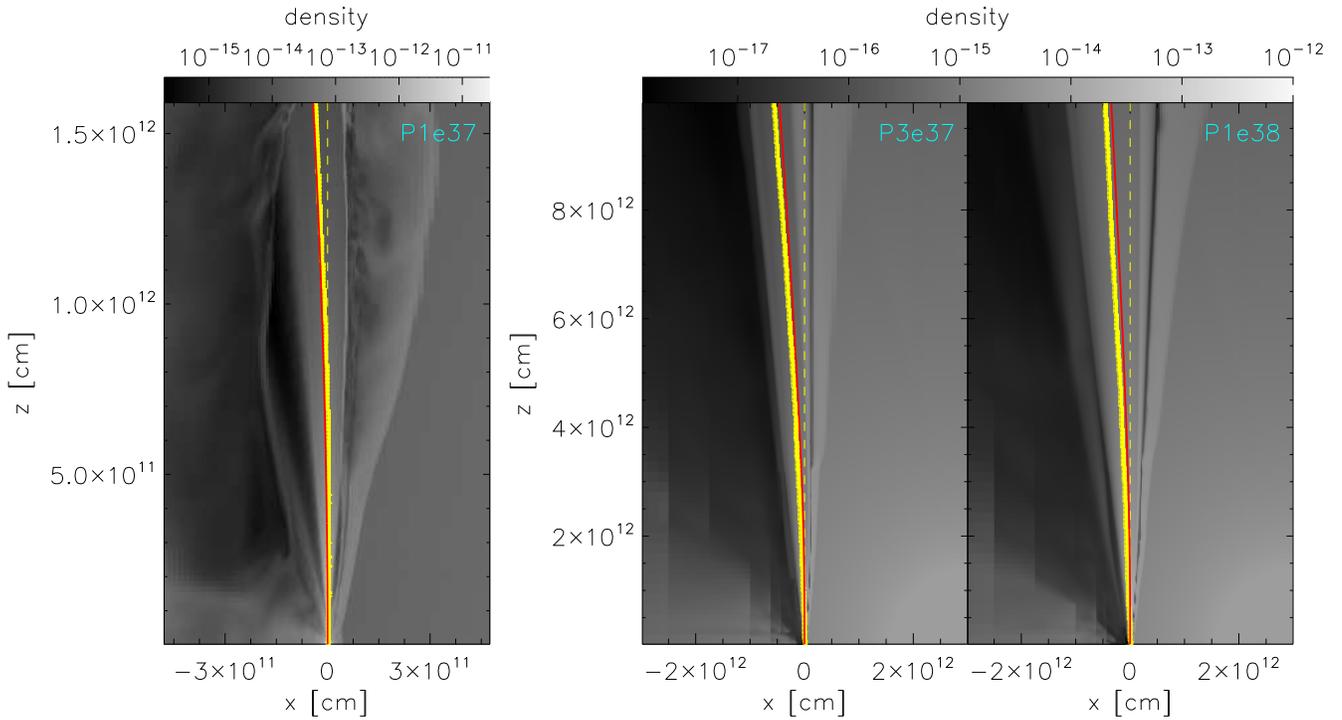}
   \caption{The density contour maps in $x$-$z$ plane for the P1e37
     (left), P3e37 (middle) and P1e38 (right) models. The P1e37 model
     is from YH15, in which the considered length along the jet is an
     order of magnitude smaller than that of the P3e37 and P1e38
     models. The vertical line up from the black hole position is
     shown by yellow dashes. The thick yellow line represents the
     identified jet centre from simulations and the thick red line
     represents the analytic approximation to the jet trajectory.}
\label{fig:bend}
\end{figure*}

\begin{figure*}
    \centering
    \includegraphics[width=15.2cm]{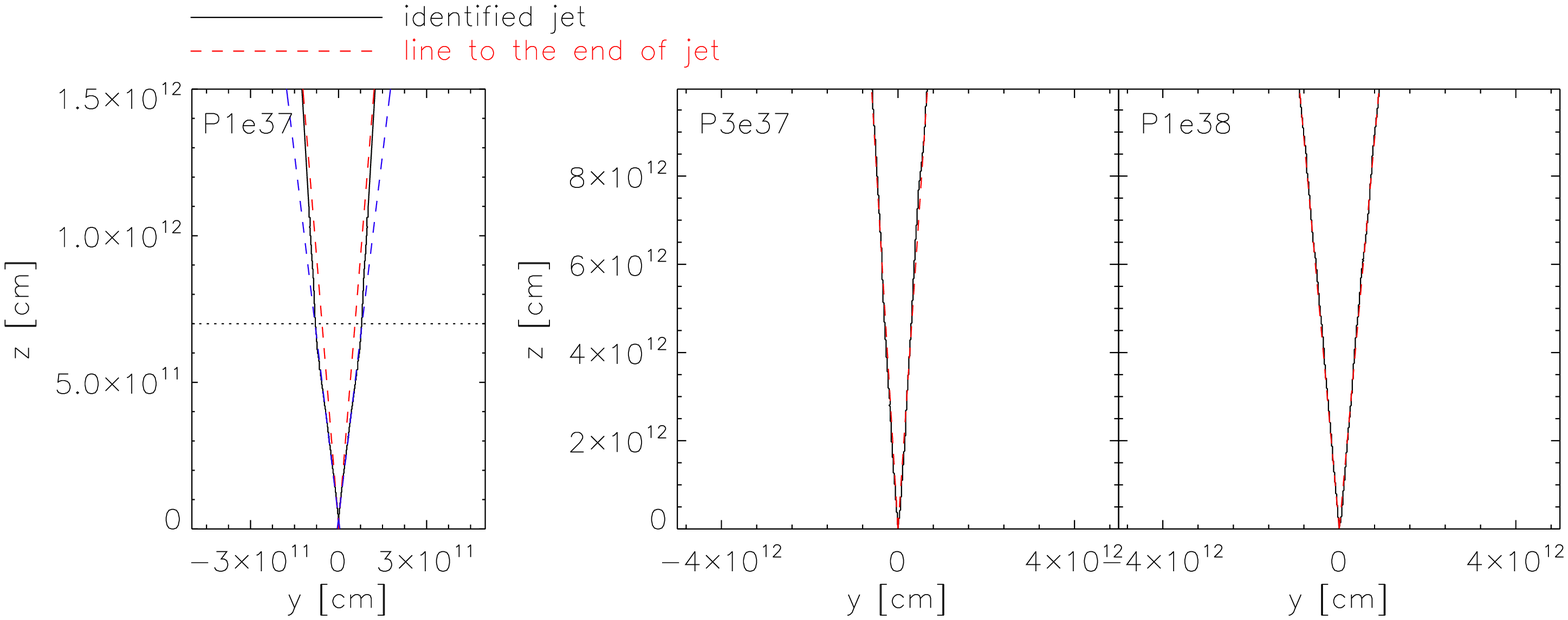}
    \caption{The jet in the $y$-direction as a function of $z$. The black
      solid line represents the jet identified from the
      simulations. The straight line between the black hole and the
      end of the jet is shown by the red dashes. In the P1e37 model,
      the blue dashes are straight lines that follow the initial
      opening angle, $\alpha_{0} \simeq 6.5\degr$, and the horizontal
      dotted line represents the location of the recollimation
      shock. }
\label{fig:jet}
\end{figure*}

\subsection{Methods}

In order to test the predictions for the formation of recollimation
shocks, we carried out 3-D hydrodynamic simulations using FLASH 3.3
grid-based adaptive-mesh hydrodynamics code \citep{Fryxell:00} for
testing the conditions in which a recollimation shock forms, using the
same method as that of YH15. In YH15, the jet was modelled by a
cylindrical nozzle with inflow-boundary conditions at the surface,
injecting a bipolar outflow with a prescribed energy, mass, and
momentum flux. This cylindrical nozzle was required to set a slow
lateral outflow from the side walls of the cylinder due to the
numerical stability.  The injected materials from the lateral outflow
interact with the stellar wind, losing the momentum and falling back
to the jet, which yields artificial mass-loading on the jet.  In the
jet model with $P_{\rm j} \leq 10^{37}\,{\rm erg\,s^{-1}}$, the amount
of the mass-loading and its effect on the jet bending are
negligible. However, as the jet kinetic power increases, the
mass-loading becomes noticeable and this effect should be taken into
consideration.  To eliminate the side effects of the artificial
outflow, we implement the jet nozzle within the lower boundary instead
of an interior grid.

Our numerical configuration is similar to that in \citet{perucho08}
and \citet{perucho10}, which showed jet-wind interaction in massive
X-ray binaries by using 2-D and 3-D simulations, respectively.  They
concluded that jet is likely bent and disrupted for jet kinetic power
$\lesssim 10^{36}\, {\rm erg\,s^{-1}}$ by the ram pressure of stellar
wind from nearby O-type star, which is also consistent with YH15.

The parameters of the jet and stellar wind are the same as in YH15,
except for the jet kinetic power. Namely, we use
$v_{\rm j}=3\times 10^{9}\,{\rm cm\,s^{-1}}$,
$v_{\rm w}=2.5\times 10^{8}\,{\rm cm\,s^{-1}}$,
$\dot{M}_{\rm w}=10^{-5}\,\msun\, {\rm yr}^{-1}$,
$a=3\times 10^{12}\,{\rm cm}$, and the Mach number of
$\mathcal{M}_{\rm j}=30$, which corresponds to one of the two
recollimation-shock cases studied by YH15. For the assumed
$v_{\rm j}$, $v_{\rm w}$, and $\dot{M}_{\rm w}$, the critical jet
power, equation (\ref{condition}), is
$P_{\rm cr}\simeq 3.0\times 10^{37}\,{\rm erg\,s^{-1}}$.

In addition to the recollimation case considered by YH15, with a
kinetic power of $P_{\rm j}= 10^{37}\,{\rm erg\,s^{-1}}$ (P1e37), we
have carried out two sets of runs with $P_{\rm j}=3\times 10^{37}$ and
$ 10^{38}\,{\rm erg\,s^{-1}}$, which we denote as P3e37 and P1e38,
respectively. In our coordinate system, the jet origin and the star
are located at $x_{\rm CO}=0$ and $x_*=a$, respectively.  A volume
rendering and contour plot of the P3e37 simulations is shown in
Fig.~\ref{fig:render} for illustration.

\subsection{Jet bending}
\label{bending}

In the presence of a stellar wind, the jet is bent by a transverse
pressure gradient that the wind drives around the jet (see
Fig.~\ref{fig:render}). The bending has to be taken into account for
identification of a recollimation shock. Therefore, we identify the
jet centre in order to be able to estimate the jet thickness in the
$y$-direction, see YH15.  Fig.\ \ref{fig:bend} shows how the jet is
deviating from its initial direction along the $z$ axis. The yellow
thick lines show the jet centre identified in the simulation data. We
ensure that the simulations have evolved for a sufficiently long time
for the initial bow shock of the jet to propagate well beyond the
region of interest, allowing for interaction directly with the
stellar wind.

For the analysis of jet-bending, we assume the jet remains conical
with a fixed jet opening angle, $\alpha_{0}$. This is in contrast to
the analytic jet model in YH15, in which the jet is confined by
pressure equilibrium between the jet and the wind (red solid line of
the P1e37 model in Fig.\ \ref{fig:bend}), and thus the jet is not
conical.  The former is appropriate for powerful jets and prior to the
formation of a strong recollimation shock, the latter is appropriate
for weaker and strongly bent jets beyond the location of the
recollimation shock. As we will show in Section \ref{recollimation} below,
the more powerful jets in our case study (the P3e37 and P1e38 models)
are not substantially recollimated, and have approximately constant
opening angle, $\alpha_{0}$, so we adopt a conical shape for the
analytic jet model.

The derivation is similar to that of YH15. Assuming that the
longitudinal jet velocity is constant and $\mathcal{M}_{\rm j} \gg 1$,
the longitudinal jet momentum per unit jet length can be calculated by
\begin{equation}
  \Phi_{\rm j}=\int{{\rm d}A_{\perp}\rho_{\rm j}\,v_{\rm
      j}}=\frac{P_{\rm j}}{v^{2}_{\rm j}},\label{Phi}
\end{equation}
where ${\rm d}A_{\perp}$ is the area perpendicular to the initial jet
direction. The transverse wind momentum per unit jet length
accumulated by the jet as a function of $z$ is
\begin{equation}\label{eq:windm}
  \Delta \Phi_{\rm w} = \frac{1}{v_{\rm j}}\int^{z}_{0}{{\rm d}z'\,
    v_{{\rm w,}x}^2 (z')\,\rho_{\rm w}(z')}\, h(z'),
\end{equation}
where $h(z)=2 z\,\tan{\alpha_{0}}$ is the jet width. Equation
(\ref{eq:windm}) can be readily integrated to
\begin{equation}
  \Delta \Phi_{\rm w} = \frac{v_{\rm w,0} \dot{M}_{\rm w}
    \tan{\alpha_{0}}}{4\upi v_{\rm
      j}}\frac{z^{2}}{a^{2}+z^{2}}.\label{DeltaPhi}
\end{equation}

In the first-order approximation, the jet bending angle, $\psi$, can
be derived by the ratio between the accumulated transverse momentum
and the longitudinal momentum as a function of $z$,
\begin{equation}\label{eq:bending}
  \psi(z)\simeq \frac{\Delta \Phi_{\rm w}}{\Phi_{\rm m,j}}  =
  \frac{v_{\rm w,0}v_{\rm j}\dot{M}_{\rm w}\tan{\alpha}}{4\upi P_{\rm
      j}}\frac{z^{2}}{a^{2}+z^{2}}.
\end{equation} 
At $z^2\gg a^2$, the bending angle reaches its asymptotic value. In
the small-angle approximation, we can derive the analytic jet
trajectory through ${\rm d}(x-x_{\rm CO})/{\rm d}z=\tan{\psi}$.
Analytic trajectories derived for the simulations are shown by the
thick red lines in Fig.\ \ref{fig:bend}. They show good agreement with
the simulation results (thick yellow lines), indicating that our
simple analytic model can describe the bending jet. The bending angle
is similar between the P3e37 and P1e38 models, because although the
P1e38 model has a jet kinetic power about 3.3 times larger than that
of the P3e37 one, the opening angle in P1e38 is also 2.4 times larger
than that of P3e37 (see Section \ref{recollimation}).

\subsection{Recollimation}
\label{recollimation}

YH15 studied the appearance of a recollimation shock at
$P_{\rm j} = 10^{37}\,{\rm erg\,s^{-1}}$, in which case the shock
takes place at $z \simeq 7\times10^{11}\,{\rm cm} \ll a$ (horizontal
dotted line in the left panel of Fig.\ \ref{fig:jet}). 

In YH15, the jet was found to initially expand freely, its lateral
expansion proceeding until the lateral ram pressure of the jet dropped
below the wind ram pressure, where a recollimation shock occurred
(independent of the initial jet opening angle and the jet Mach
number). Beyond the shock, the jet evolved adiabatically in pressure
equilibrium between the jet and the wind bow shock.

However, with increasing jet power, the shock takes place at
increasingly larger $z$ or does not occur at all, as found
analytically in Section \ref{formation}. To check for the presence of
a recollimation shock, we measured the jet width facing the star,
i.e., along the $y$-direction. We measured the width along the
identified jet centre (yellow thick line in Fig.\ \ref{fig:bend}). The
black solid lines in Fig.\ \ref{fig:jet} show that the measured jet
width increases almost linearly, and no recollimation shock is
apparent. The opening angles are $\alpha({\rm P3e37})\simeq 3.4\degr$
and $\alpha({\rm P1e38})\simeq 8\degr$.

\begin{figure}
    \centering
    \includegraphics[width=\columnwidth]{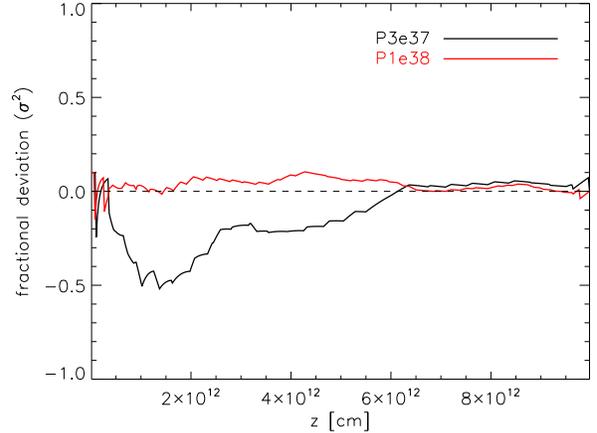}
    \caption{The squared fractional deviation between the jet edge measured in
      simulations in the $y$-direction and the straight line
      connecting the black hole and the jet end, given by $\sigma^2$
      defined by equation (\ref{sigma}). The black and red colours
      correspond to the P3e37 and P1e38 models, respectively. The
      horizontal dashed line indicates no deviation.}
\label{fig:dev}
\end{figure}

In order to determine whether the contact discontinuity is measurably
misaligned with the initial jet opening angle, we compare the jet
trajectories with straight lines connecting the black hole with the
edge of the jet at the upper limit of the analysis box (red dashes in
Fig.\ \ref{fig:jet}). Then we calculate the (squared) fractional deviation as a
function of $z$,
\begin{equation}
\sigma^2(z) \equiv [y_{\rm j0}^{2}(z)-y_{\rm j}^{2}(z)]/y_{\rm j0}^{2}(z),
\label{sigma}
\end{equation}
where $y_{\rm j}$ and $y_{\rm j0}$ are the positions of jet edge for
the simulations and the straight line, respectively.  Fig.\
\ref{fig:dev} shows that for P1e38, the identified jet boundary is in
good agreement with the straight line, indicating a conical shape of
jet and the absence of a recollimation shock.  

The jet in P3e37 also has an overall almost conical shape, but there
is substantial negative deviation from this shape at
$z \la 3\times 10^{12}\,{\rm cm}$. This shows that a surface shock
does take place around $z\sim a$, confirming the analytic prediction
of Section \ref{formation} for the shock location around the
separation at $P_{\rm j}\gtrsim P_{\rm cr}$. The jet at $z\la a$ has then
an opening angle $\alpha >3.4\degr$.

These results are in an good agreement with the predictions of
equation (\ref{condition}) for the wind properties that we have used
in this work.  According to equation (\ref{condition}), the critical
jet power, above which there should be no strong recollimation shocks,
equals $P_{\rm j}\simeq 3.0\times 10^{37}\,{\rm erg\,s^{-1}}$. From
our simulations, there is a strong shock at
$P_{\rm j}= 10^{37}\,{\rm erg\,s^{-1}}$, a mild shock at
$P_{\rm j}= 3\times 10^{37}\,{\rm erg\,s^{-1}}$, and and no
measureable change in opening angle for
$P_{\rm j}= 10^{38}\,{\rm erg\,s^{-1}}$.

While the analysis presented in this paper focuses on the shock formed
by the interaction of the leading edge of the jet with the stellar
wind bow shock, the simulations allow a description of the down-wind
half of the jet dynamics as well. As can be seen in
Fig.~\ref{fig:render}, the thermal pressure of the bow shock drops to
low values in the shadow of the jet, leading to the formation of an
expansion fan that accelerates the jet and wind fluid into the
evacuated region behind the jet. In some instances, this very
low-density material may form a shock upon convergence on the $x$-$z$
plane behind the jet, which is the reason for the appearance of denser
gas in the wake of the jet in Fig.~\ref{fig:bend}.

\section{Jets in Cyg X-1 and Cyg X-3}
\label{cyg}

We can check the implications of the above results for two high-mass
X-ray binaries with jets. Cyg X-1 contains a black hole with the mass
of $\simeq 16\msun$ \citep{orosz11,ziolkowski14}. The mass of the
donor is disputed, \citet{ziolkowski14} estimates it as
$\simeq 27\msun$. Then, the separation equals $3.2\times 10^{12}$
cm. Cyg X-1 has a steady compact jet in the hard state
\citep{stirling01}.

The current best estimates of the relevant parameters of Cyg X-1 are the mass-loss rate in the hard state of $\dot M_{\rm w}\simeq 2.6\times 10^{-6}\msun$ yr$^{-1}$ \citep{gies03}, its asymptotic velocity of $v_{\rm w}\simeq 1.6\times 10^8$ cm s$^{-1}$ \citep{gb86} and the relatively uncertain jet velocity of $v_{\rm j}\simeq 0.6 c$ \citep*{stirling01,gleissner04,mbf09}. These parameters correspond to the critical power of $P_{\rm cr}\simeq 3\times 10^{37}$ erg s$^{-1}$. This is within the estimated $P_{\rm j}\simeq (1$--$7)\times 10^{37}$ erg s$^{-1}$ (for the jet+counterjet) from the optical nebula presumably powered by the jet (\citealt{sell15}; see also \citealt{gallo05,russell07}). Thus, a recollimation shock can occur in the hard-state jet of Cyg X-1. The location of the shock is given by equation (\ref{solution}). At $P_{\rm j}=10^{37}$ erg s$^{-1}$, the shock takes place at $z=0.32 a$. The jet power of Cyg X-1 has also been estimated in several theoretical models of \citet*{zls12}, \citet*{mzc13}, \citet{z14b} and YH15. For almost all of the model power was $\ll P_{\rm cr}$ (the only exception was one model in \citealt{mzc13}, which those authors themselves considered unlikely).

Although it is likely that Cyg X-3 also contains a black hole, the
current solution for its mass also allows the presence of a neutron
star \citep*{zmb13}. We adopt
$\dot M_{\rm w}\simeq 7\times 10^{-6}\msun$ yr$^{-1}$ (see
\citealt{zmb13} and references therein),
$v_{\rm w}\simeq 1.6\times 10^8$ cm s$^{-1}$ \citep{vk96},
$v_{\rm j}\simeq 0.5 c$ \citep{z12,dch10} and $a=2.4\times 10^{11}$ cm
(assuming the total mass of $13\msun$, \citealt{zmb13}). We then
obtain $P_{\rm cr}\simeq 6.6\times 10^{37}$ erg s$^{-1}$. This is
similar to, or somewhat below, the power estimate of \citet{z12}
during periods of strong \g-ray emission in the soft state
\citep{fermi,agile}, based on the \g-ray luminosity.

Thus, a strong recollimation shock may occur in the
hard-state jet of Cyg X-1 and the soft state of Cyg X-3 during periods
of strong \g-ray emission. Since $P_{\rm j}\sim P_{\rm cr}$ in both
cases, the shock location is close to the maximum possible position of
$z\sim a$.  This, in fact, agrees with the analyses of \citet{dch10}
and \citet{z12} for Cyg X-3, based on modelling of the \g-ray
emission.

In Cyg X-1, the formation of a recollimation shock at $z\la a$ may
thus explain an apparent departure of the structure of the jet from
the standard model of \citet{bk79}, see \citet{heinz06}, \citet{zdz12}
and \citet{zls12}. Namely, the observed resolved fraction of the
8.4-GHz jet emission \citep{stirling01} implies the location of the
onset of the optically-thin synchrotron radio emission at
$z\sim 10^{14}$ cm \citep{heinz06}, which appears not compatible with
the observed orbital modulation of the 15-GHz emission having a depth
of $\simeq 30$ per cent \citep{zdz12}. This modulation results from
free-free absorption in the wind, and its large depth requires a
substantial difference between the paths through the wind between the
superior and inferior conjunctions, which, in turn, implies that a
large part of the radio emission originates from $z\la a$
\citep{zdz12}. A recollimation shock at $z\la a$ can lead to strong
dissipation and radio emission (spatially unresolved) on top of the
remainder of the jet emission, explaining the above discrepancy.

On the other hand, we note that Cyg X-1 in its hard state is much less radio loud than Cyg X-3 in its hard state. For the radio loudness defined as the ratio of $\nu L_\nu$ at 15 GHz to the bolometric luminosity (dominated by X-rays), both in units of the Eddington luminosity, this striking difference is illustrated in fig.\ 7 of \citet*{z16}. This ratio is $\sim 3\times 10^{-8}$ and $\sim 10^{-6}$ in Cyg X-1 and Cyg X-3, respectively. This difference by a factor of $\sim$30 could be explained by the shock in the former and the latter being weak and strong, respectively.

Finally, we stress that bright radio jets are common also in low-mass
X-ray binaries, e.g., \citet{fender06}, where no stellar wind-jet
interaction occurs. Thus, the formation of a recollimation shock,
either strong or weak, is clearly not a condition for a jet radio
emission; however, it is likely to modify the radio emissivity profile
with respect to that corresponding to the absence of such a shock.

\section{Conclusions}

We have derived an analytical criterion for the existence of
recollimation shocks in jets in high-mass X-ray binaries. We have
tested it with 3-D numerical simulations and have found an good
agreement with the analytic predictions. According to this criterion,
a recollimation shock in jets of high-mass X-ray binaries can exist
only below the critical jet power proportional to the mass-loss rate,
given by equation (\ref{condition}). Below that critical jet power,
the position of the shock is given by equation (\ref{solution}). Both
the condition for shock formation and the shock location are
independent of the initial jet opening angle.  Above the critical
power, a highly oblique surface shock may still form to raise the
jet's internal pressure to the bow shock pressure, but such a shock
will only change the opening angle by a small amount and will only
occupy a small fraction of the jet's volume.

We have applied our results to the high-mass binaries Cyg X-1 and Cyg
X-3. The presence of a recollimation shock is likely in both of those
objects. In the latter system, we can identify the position of the
shock with the region of \g-ray emission. The large difference in the
radio loudness between those objects can be explained by the presence
of a weak and strong shock in Cyg X-1 and Cyg X-3, respectively.

\section*{ACKNOWLEDGMENTS}

We thank Marek Sikora for valuable discussions, and Guillaume Dubus and the referee for valuable comments. AAZ has been supported in part by the Polish NCN grants 2012/04/M/ST9/00780 and 2013/10/M/ST9/00729. DSY and SH acknowledge support from the National Science Foundation through grant AST 0908690. This work used the Extreme Science and Engineering Discovery Environment (XSEDE), which is supported by National Science Foundation grant number ACI-1053575.

\label{lastpage}

\end{document}